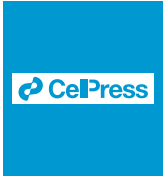
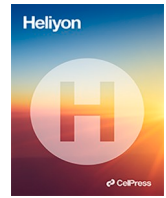

# Photovoltaic energy sharing: Implementation and tests on a real collective self-consumption system

Haritza Camblong [a,d], Octavian Curea [b], Juanjo Ugartemendia [c,*], Zina Boussaada [b], Iban Lizarralde [b], Garazi Etxegarai [a,b]

[a] *Department of Systems Engineering & Control, Faculty of Engeneering of Gipuzkoa, University of the Basque Country (UPV-EHU), Europa Plaza 1, E-20018, Donostia, Spain*
[b] *University of Bordeaux, ESTIA Institute of Technology, Technopole Izarbel, 64210, Bidart, France*
[c] *Department of Electrical Engineering, Faculty of Engineering of Gipuzkoa, University of the Basque Country (UPV-EHU), Europa Plaza 1, E-20018, Donostia, Spain*
[d] *Department of Electrical and Electronic Engineering, Auckland University of Technology, Auckland, 1010, New Zealand*

A B S T R A C T

This research study analyses different types of photovoltaic (PV) energy sharing in a collective self-consumption (CSC) real-case in the Izarbel technological park in France. The analysis is carried out above all from the point of view of the self-consumption rate (SCR) and the savings.

After explaining the emergence of the self-consumption concept for the integration of renewable energies, the study case is described. The PV energy is produced in ESTIA1 building and consumed in ESTIA1, 2 and 4 buildings. The main IoT components used to implement the CSC are smart meters and the Tecsol TICs; devices based on the LoRa protocol to retrieve production and consumption data. Then, the characteristics of PV energy sharing in France are explained, in particular the three possible types of energy sharing/allocation (static, dynamic by default and customised dynamic) and the structure of the electricity bill. Finally, the three types of sharing are compared in four scenarios (without and with a data centre, for low and high solar radiation). The results show that the dynamic allocations lead to increases of the SCR and that the customised dynamic sharing increases savings.

## 1. Introduction

Once again, the latest Climate Change Convention, COP 27, highlighted the urgent need to take action to mitigate the consequences of climate change [1]. With this in mind, the Intergovernmental Panel on Climate Change recently published the third part of the Sixth Assessment Report, Climate Change 2022: Mitigation of Climate Change [2]. In this report, decentralised renewable energy is seen as one of the keys that will contribute to climate change mitigation, while creating economic opportunities.

Micro-grid topology seems to be an interesting solution to decentralised renewable energy sources. A micro-grid is a local electricity network that facilitates the integration of distributed energy sources [3]. Several works show solutions for integrating different sources, such as microturbines, photovoltaic (PV) panels, fuel cells or wind turbines. In Refs. [4,5] works, the scheduling of all of them in complex micro-grids is analysed.

Another relatively new concept related to the integration of local renewable energies is the self-consumption. Unlike micro-grids, the self-consumption concept does not require frequency and voltage regulation, and is therefore simpler to manage. The self-consumption can be individual or collective, based mainly on renewable energy [6]. Moreover, it can be considered in the context of Directive (EU) 2018/2001 on energy communities (EC) [7–10]. Whether individual or collective, the main objectives of self-consumption are to maximize the self-consumption rate and the economic benefits [11,12]. In the case of individual self-consumption, these objectives largely depend on the correct sizing of the electricity production system, which is generally based on PV panels [13,14]. A good management of flexible loads (FL) can also improve the self-consumption rate and the economic benefits [15,16].

In the case of collective self-consumption (CSC), in addition to the production sizing and the management of FL, the way in what the electricity produced is shared between consumers has a significant impact on the objectives mentioned [17].

The design and implementation of individual self-consumption installations can be considered mature, given that this is it the type of self-consumption used in recent decades [18,19]. CSC concept is more innovative. That is why there are not many publications on this topic in the scientific literature. However, articles on the regulatory framework in different countries are available [20–26], and some of them deal with the type of energy sharing permitted in each country. In general, three types of allocation are regulated: static (pre-defined), dynamic (post-defined) and variable (pre-defined).

Research studies that address the issue of sharing/allocation are mainly based on theoretical simulations or use real data, but are not applied to real installations [27]. In some of them [17,28], the sharing criterion is based on the fairness. There are not many studies comparing the different types of energy sharing. In Ref. [29], the authors compare static and dynamic allocation in an industrial park of the Autonomous Basque Community, in the north of Spain. The authors conclude that economic benefits are slightly higher when dynamic allocation is applied. In Ref. [26] the authors propose a dynamic allocation coefficient to allow members of the community to share their surplus with other members, thereby increasing the self-consumption rate.

France, where the principle of collective self-consumption was introduced in 2016 [30], appears to have some of the most favourable legislations for CSC in Europe [9], particularly in terms of the size of the CSC area and the types of sharing. In French legislation, CSC is defined as the sharing of local production directly with consumers via the public distribution network. All types of energy are permitted, not just PV, but up to a maximum of 3 MW. Many different producers and consumers can participate in a 2 km diameter area. The power station may be owned by the consumers themselves or operated by a third party who sells the electricity to the consumers.

As mentioned above, to date, little work has been published on CSC, and even less on the specific case of French regulation. In Ref. [31], the sharing/allocation of the overall community generation is denoted as keys of repartition (KoR). These keys are co-efficients designed on the basis of different criteria (e.g. equal allocation, sharing proportional to consumption, proportional to production, etc.). Some of these keys are static and others dynamic. The simulation results, based on real data, show that when consumers are grouped together in a CSC community, a reduction of 11.5 % in the community's overall bill and a reduction of between 11 % and 19 % in individual bills can be achieved.

The original research work described in this article is based on the research study carried out as part of the EKATE project funded under the 3rd call of the INTERREG V-A Spain-France-Andorra programme (POCTEFA 2014–2020) [32]. The EKATE project aimed to develop efficient energy sharing and management systems for CSC PV installations, using Blockchain and Internet of Things (IoT) technologies. One of the two pilot actions of the project was located in the Izarbel technology park in Bidart, France. This pilot project involved three tertiary sector buildings.

Considering the above study on the current scientific literature, **the main contribution of the research study presented in this**

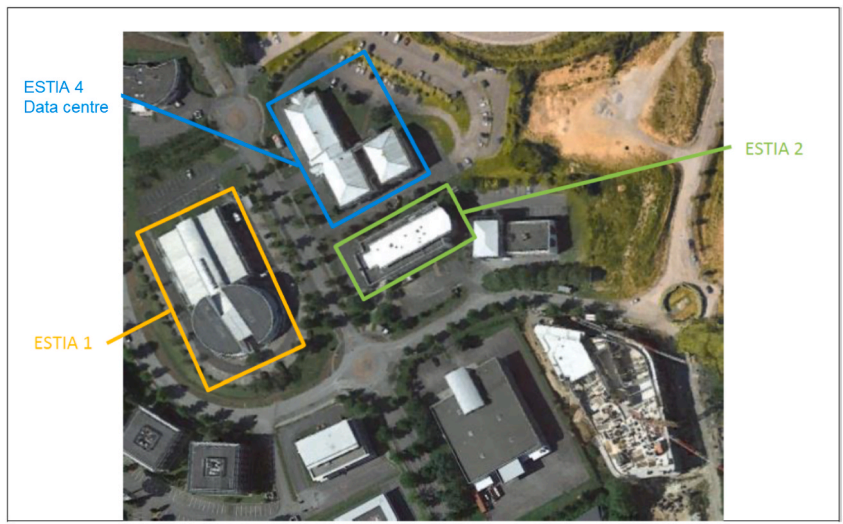

**Fig. 1.** The three buildings involved in the CSC of Izarbel technology park.

paper concerns the comparison on a real CSC application of the three possible energy-sharing types regulated in France, using the self-consumption rate and the economic benefits as comparison criteria. The application of a customised dynamic allocation to a real system can also be highlighted.

The paper is divided into five sections. Section 2 describes the studied CSC case. Section 3 depicts the characteristics of the allocation of the PV energy between the three buildings involved in the CSC. Then, section 4 shows and analyses the results obtained by applying the different energy allocation types. Finally, the conclusions related to this research study and some perspectives are given in section 5.

## 2. Study case

### 2.1. General characteristics of the Izarbel CSC study case

The PV energy CSC demonstrator in the Izarbel technology park is composed of three buildings, ESTIA1, ESTIA2 and ESTIA4 (business centre, which contains in particular a data centre). All three buildings (see Fig. 1) are managed by the ESTIA Institute of Technology engineering school.

The study on the installation of PV panels and the CSC carried out by Tecsol Company (member of the consortium of EKATE) provided for the installation of a total of 286 kWp of PV panels in these three buildings.

In a first step, only ESTIA1, and especially the most profitable area of this building (see the rounded roof in Fig. 1) would be equipped with PV panels. This area faces south-east and has a slope of 20 %. A total of 149 kWp would be installed in the roof of ESTIA1. Thus, after this first step, the PV energy would be produced in ESTIA1 and consumed in the three buildings mentioned above. The rest of the PV panels would be installed at a later stage.

As the above-mentioned PV were not yet installed during the EKATE project, the existing 5.6 kWp PV installation at ESTIA1 (visible in the roof in Fig. 1, installed in 2005) was used to emulate the 149 kWp PV panels to be installed shortly. For the emulation, the measurement of the electricity production of the 5.6 kWp PV panels, obtained via the associated production meter, was multiplied by a gain. The gain was calculated by dividing the maximum power that the new panels would produce according to the PV Syst software [33] used by Tecsol in its study, i.e. 118.5 kW, by the maximum power produced in 2021 by the current panels, i.e. 4.65 kW. The resulting gain value was 25.48. The fact that the new PV panels will have the same location and orientation as those already installed is a solid guarantee of the quality of the emulation.

A specific feature of the data centre located at ESTIA4 is that the electricity consumption of the related servers varies very little throughout the 24 h a day, 7 days a week. Thus, with the inclusion of the data centre in the legal entity responsible of the CSC of Izarbel (*Personne Morale Organisatrice*, PMO, in French), a permanent consumption of more than 100 kW would be ensured. This fact would allow self-consumption of excess production during weekends and holidays, especially during the summer weeks when students are on holiday, i.e. when the activity at ESTIA1 and ESTIA2 is minimal.

As the 149 kWp PV panels were not yet installed during the project, the administrative procedures related to the creation of the legal entity organising the CSC was not completed. Therefore, the data centre participation in the CSC was not yet confirmed.

### 2.2. Developed and installed IoT

#### 2.2.1. Global architecture of the IoT developed in EKATE project

This sub-section describes the IoT required to implement and operate in real-time the energy sharing and management. Fig. 2 depicts the general architecture of the Izarbel IoT infrastructure. The core of the architecture is a Raspberry Pi single-board computer.

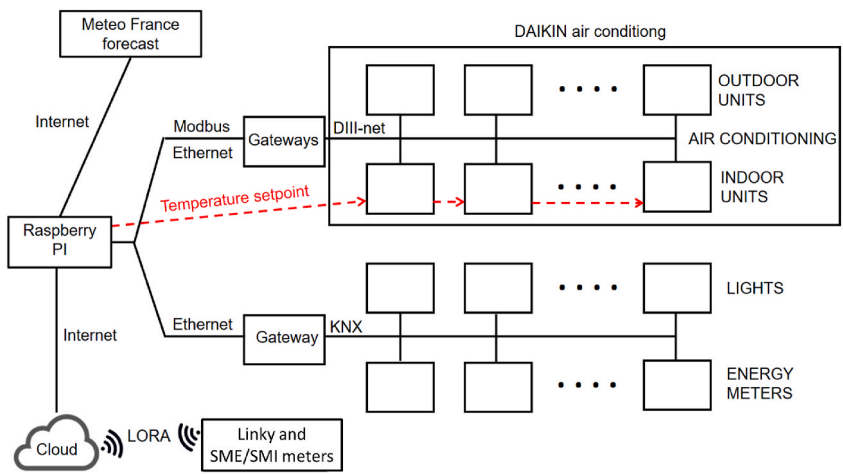

**Fig. 2.** IoT infrastructure block diagram.

In particular, connections with the following components can be highlighted:

- The Météo France website.
- The indoor and outdoor units of the ESTIA2 Daikin heat, ventilation and air conditioning system, via a Modbus TCP/IP bus and a gateway linking to the DIII-net bus.
- The lighting and counters of the HP via Ethernet (TCP) and a gateway linking to the KNX bus.
- The cloud and the LoRa network (access to Linky and SME meters).
- The weather station installed on the ESTIA2 rooftop.

The following subsection presents the most important IoT components used to manage the sharing of the PV energy between the three aforementioned buildings.

*2.2.2. Smart meters and Tecsol TIC*

Meters data are retrieved thanks to the electronic devices designed by Tecsol [34]. Fig. 3 shows the Tecsol TICs connected to the meters of the three buildings involved in the PMO. This Figure represents the connection between the meters and the cloud shown in Fig. 2.

The Tecsol TIC antenna enables communication over the LoRa network with the LoRaWAN (Long Range Wide Area Network) protocol [35,36]. This is a radio communication protocol based on LoRa technology.

In the Izarbel pilot project, Tecsol TICs retrieve the data from the smart meters and send it to the Tecsol server via the LoRa network. Sunchain, the start-up linked to Tecsol [37] that manages CSC static and customised dynamic sharing (see Table 2) according to the choices carried out by the PMO, can retrieve the data needed for the KoR calculation through an application-programming interface (API).

Table 1 lists the information of the Izarbel communicating meters involved in the CSC, pending whether or not the data centre will integrate the PMO. There are 4 m in total. Two of them are in ESTIA1: a consumption meter and a meter measuring the production of the PV panels installed at ESTIA1. ESTIA2 contains a single consumption meter. As for ESTIA4, as a business centre, it contains several meters. For the moment, only one participates in the CSC, the consumption meter for some offices. If the data centre was included in the PMO, it would be associated to this meter.

The accessible data and their resolution differ depending on the type of meter. In France, SME/SMI meters are used when the contacted power is higher than 36 kVA. These meters provide, among other things, the rate of active energy withdrawn with a resolution of 1 kWh and the average active power during 10 min in extraction with a resolution of 1 kW. Linky meters are used when the contracted power is lower than 36 kVA. They provide the total active energy injected (in production) or withdrawn (in consumption), with a resolution of 1 Wh, and the instantaneous apparent power injected or withdrawn in VA.

Linky meters therefore offer much better resolution than SME/SMI meters. The low resolution of SME/SMI meters is a drawback, especially at times when the consumption of a building equipped with this type of meter is low.

## 3. Characteristics of the sharing of the PV energy between the three buildings involved in the CSC

*3.1. CSC energy sharing in France*

The EKATE project focused on the deployment of an innovative scheme for the use of the PV energy produced, i.e. collective or shared self-consumption. In this concept, the energy produced is allocated to the various consumption points via the public network. This scheme is made possible by Article 119 of the Law for Energy Transition [38], confirmed by Order 2016-1019 [39], ratified by

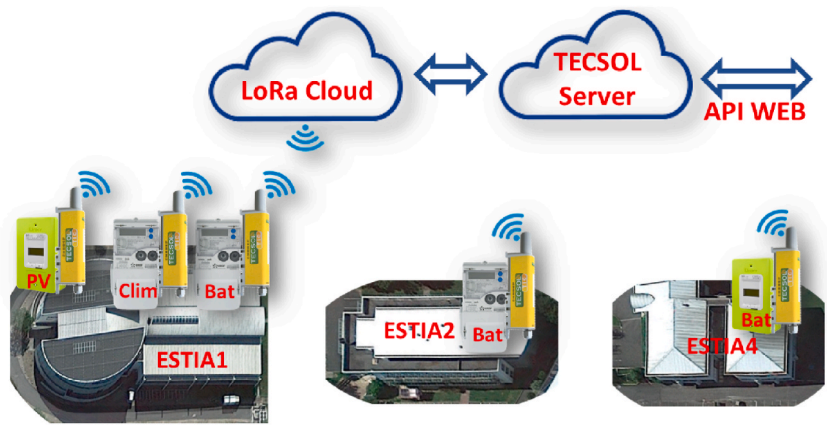

**Fig. 3.** Tescol TICs connected to the meters in the three buildings.

**Table 1**
Meters implicated in the Izarbel CSC.

| Building | Meter type | Observation | Data retrieved | Resolution | Definition |
|---|---|---|---|---|---|
| ESTIA1 | SME/SMI | ESTIA1 consumption | P+ | kWh | Active energy extraction index |
|  |  |  | 10min extracted power | kW | Average active power over 10 min in extraction |
| ESTIA1 | Linky | ESTIA1 Prod PV | P- | Wh | Total injected active energy |
|  |  | 5.6 kWp for sale in full | 10 min injected power | VA | Instantaneous injected apparent power |
| ESTIA2 | SME/SMI | ESTIA2 consumption | P+ | kWh | Active energy extraction index |
|  |  |  | 10min extracted power | kW | Average active power over 10 min in extraction |
| ESTIA4 | Linky | ESTIA4 consumption of some offices | P+ | Wh | Total extracted active energy |
|  |  |  | 10min extracted power | VA | Instantaneous extracted apparent power |

Law No. 2017-227 of 24 February [40] and implemented by Decree No. 2017-676 of 28 April [41].

Physically, the electricity produced and not consumed on site is injected into the grid, but rather than being valued through a purchase contract, it is allocated to the needs of the site through the modification of information obtained from the site's consumption meters.

The regulations designate the CSC PMO (a single entity with regard to the public grid operator) as responsible for allocating the production to the various consumption points. The dispatch is carried out and accounted for with a time step of 30 min.

As illustrated in Table 2, this energy allocation can be carried out statically, dynamically by default or dynamically (customised).

In the static case, the KoRs are constant. In the default dynamic case, the KoRs are variable and automatically calculated by Enedis French DSO, in proportion to the consumption of each participant. Finally, in the third and last case, the customised dynamic allocation, the sharing is carried out according to precise and well thought-out rules established by the CSC PMO. In EKATE, this allocation was managed by Sunchain.

The CSC schemes are completed with the intervention of a third party for the management of the dynamic KoRs provided to Enedis. In EKATE, this third party was the company Sunchain, as above mentioned, a start-up from the company Tecsol. A patent was filed by Sunchain for the sharing of flows: it provides the original KoRs between energy producers and consumers.

The diagram in Fig. 4 shows the information flow between the different actors, in order to meet the requirements of the CSC model as described by the current set of regulations in France.

Sunchain has an experimental agreement with Enedis, which enables the development of a data exchange interface between Enedis and the CSC projects.

*3.2. KoR management through blockchain*

The quantities of energy measured at fine time steps represent a large volume of data, which it is important to store, but also to certify and secure. Energy consumption, in particular, falls within the scope of personal data and is subject to regulations on the protection of privacy. Blockchain technology was chosen for this issue because of the private nature of the group created by the participants in a CSC operation, which operates on the principle of a distributed computer consensus of the Proof of Authority type.

The information recorded in the blockchain is:

- The identification of the counting point through its private/public key.
- A timestamp.
- Energy data (production and consumption) at the chosen time step.

The data is collected at the meter level in a secure and tamper-proof way, before being integrated into the blockchain in an encrypted way. The KoRs are calculated according to the equations corresponding to the sharing wishes of the organising entity.

*3.3. Self-consumption rate and economic issues*

As mentioned above, the two main objectives of a CSC system are to maximize the self-consumption rate and the economic benefits. Regarding the self-consumption rate (SCR), it is defined [42] in Eq. (1):

$$SCR = \frac{\sum_{1}^{48} E_{B_{SC}}(k)}{\sum_{1}^{48} E_{PV}(k)}, \tag{1}$$

where $E_{B_{SC}}(k)$ represents the self-consumed energy in the three buildings in each calculation period $k$ as shown in Eq. (2):

$$E_{B_{SC}}(k) = E_{B1_{SC}}(k) + E_{B2_{SC}}(k) + E_{B4_{SC}}(k), \tag{2}$$

where, $E_{B1_{SC}}(k)$, $E_{B2_{SC}}(k)$ and $E_{B4_{SC}}(k)$ are ESTIA1, ESTIA2 and ESTIA4 buildings self consumption amount in each period, respectively. $E_{PV}(k)$ represents the PV energy produced in each considered $k$ period. In France, the SCR is calculated every 30 min [38].

**Table 2**
Information architecture related to KoR.

| Sharing Type | Description | Use cases | Advantages | Disadvantages |
| --- | --- | --- | --- | --- |
| STATIC | Constant KoR at each 30min time step (defined in advance by the PMO) | Example: KoRs assigned by the PMO up to the amount of financing provided by each participant | Ease of use for PMO | Sharing of production not optimal. High risk of having surplus production |
| DEFAULT DYNAMIC | Variable KoRs at each 30min time step, automatically calculated by Enedis in proportion to the consumption of each participant | "Local community" model (all participants are community-owned sites) | Automatic optimization. No values to communicate to Enedis | No differentiation between consumers due to specific features of the operation |
| DYNAMIC (customised) | Variable KoR at each 30min time step (defined by the PMO a posteriori) | Search for a better fit than default. Need for prioritization between participants | Possible improvement by hand of the PMO. Possible prioritization of production assignment | Every month, need to communicate to Enedis the value of the KoRs |

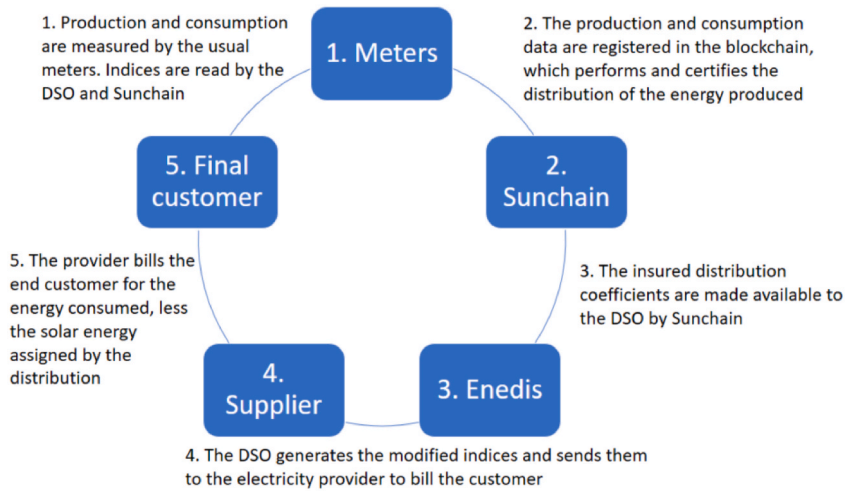

**Fig. 4.** Information architecture related to delivery and the Blockchain.

As far as the economic part is concerned, in EKATE, since the objective was to compare the relative savings under each scenario and allocation type, the PV installation investment was not considered in the saving calculation. The relative savings were calculated according to the self-consumed energy in each building and the PV surplus injected to the grid.

Three tariff-components were considered: the electricity part, the tariff for use of the public electricity grid (grid tariff) and taxes. If part of the production had to use the public distribution network to be consumed elsewhere, as in ESTIA2 and 4, the grid tariff was applied. Moreover, if a different company was consuming part of the production, as in ESTIA4, taxes were also applied.

Concerning the electricity rates, they had these values when the sharing tests were performed: the tariff of ESTIA1 and ESTIA2 was $t_{12} = 0.13$ €/kWh and that of ESTIA4 (in both scenario) was $t_4 = 0.11$ €/kWh. The overall surplus that was fed back into the grid was paid 0.06 €/kWh by EDF OA (*Obligation d'Achat*, purchase obligation or feed in tariff).

Therefore, the savings $S_{B1}$, $S_{B2}$, and $S_{B4}$ corresponding to the self-consumed energy by each building and for one day were calculated respectively as shown in Eq. (3), Eq. (4) and Eq. (5):

$$S_{B1} = \left( \sum_{k=1}^{48} E_{B1_{SC}}(k) \right) \times t_{12} \times \left( 1 + \frac{c_g + c_t}{100} \right) \quad (3)$$

$$S_{B2} = \left( \sum_{k=1}^{48} E_{B2_{SC}}(k) \right) \times t_{12} \times \left( 1 + \frac{c_t}{100} \right) \quad (4)$$

$$S_{B4} = \left( \sum_{k=1}^{48} E_{B4_{SC}}(k) \right) \times t_4, \quad (5)$$

where $t_{12}$ and $t_4$ represents the tariffs of ESTIA1, 2 and ESTIA4 respectively and $c_g$ is the variable parts of the grid tariff (called TURPE in France). This tariff is calculated following the information that is published by the DSO (ENEDIS) [43]. $c_t$ is the variable part of the other taxes that are not related to the grid (CSPE, currently TICFE in France) [44,45].

The fact is that in France, the building that is equipped with PV panels do not pay the CSPE nor the TURPE. As the PV panels were installed in ESTIA 1, this building did not pay these taxes. Therefore the savings increased compared with ESTIA4 were 66 % (28 % due to the not payment of the TURPE, and 38 % due to the not payment of the CSPE).

Concerning the ESTIA 2 building. This building had to pay the TURPE (as the grid was being used). Therefore, the 28 % due to the non-payment of the TURPE were not applied in the savings. Only the 38 % due to the non-payment of the CSPE were applied (this tax was not paid because the owner of ESTIA1 and ESTIA2 is the same).

These considerations can be summarized as follows in the savings calculations:

- As above mentioned, an average cost, $t_4$, of 0.11 €/kwh was considered for all the companies of ESTIA4.
- ESTIA2 savings were increased by 38 % due to the variable part of the taxes, $c_t$ (compared with ESTIA4).
- ESTIA1 savings were increased by 66 % (28 % due to the variable part of the grid tariff, $c_g$, and 38 % due to the variable part of the taxes, $c_t$).

Regarding the savings related to the overall surplus fed back to the grid, $S_{FB_G}$, they were calculated using Eq. (6):

$$S_{FB_G} = (E_{PV}(k) - E_{B1_{SC}}(k) - E_{B2_{SC}}(k) - E_{B4_{SC}}(k)) \times t_{fb}, \quad (6)$$

where $t_{fb}$ is the price for overall surplus fed back to the grid.

So, the total savings for the CSC were calculated using Eq. (7):

$$S_T = S_{B1} + S_{B2} + S_{B4} + S_{FB_g} \quad (7)$$

*3.4. Study methodology and definition of allocation percentages and rules*

As mentioned above, a decision on the inclusion of the data centre of ESTIA4 in the PMO was yet not taken during the EKATE project. Therefore, two scenario were considered in this research study:

1. The situation during the EKATE project, i.e. sharing between ESTIA1, ESTIA2 and some offices of ESTIA4.
2. The situation of the Tecsol study, i.e. with the consumption at ESTIA1, ESTIA2, some offices of ESTIA4 and the data centre. These two consumptions were considered linked to a unique meter of ESTIA4 that had a contracted power of 120 kVA. 100 kW of power were added to the power consumption measured in the meter of ESTIA4 in order to represent the consumption of the data centre (see Table 1).

In order to carry out a comparison between the different types of sharing in the two above-mentioned scenarios, the allocation percentages and rules had to be defined respectively for the static allocation and the customised dynamic allocation. The default dynamic allocation was calculated in proportion to the consumption of each considered meter, as managed by the French DSO Enedis. It was calculated after each 30 min period as shown in Eq. (8):

$$\text{for } i = 1, 2, 4$$
$$\text{if } (E_{B1_C}(k) + E_{B2_C}(k) + E_{B4_C}(k)) < E_{PV}(k) \text{ then } E_{Bi_{SC}}(k) = E_{Bi_C}(k) \quad (8)$$
$$\text{else } E_{Bi_{SC}}(k) = \frac{E_{Bi_C}(k)}{E_{B1_C}(k) + E_{B2_C}(k) + E_{B4_C}(k)} \times E_{PV}(k),$$

where $E_{B1_C}$, $E_{B2_C}$ and $E_{B4_C}$ represent the consumption of ESTIA1, ESTIA2 and ESTIA4 buildings respectively and where $i$ index represents each building.

Regarding the static sharing, the allocation percentages were defined according to the consumption related to the mentioned three buildings meters during one year, from 17/10/2020 to 16/10/2021. The obtained allocation percentages are given in Table 3.

The self-consumed energy in each building for the first scenario was calculated using Eq. (9):

$$E_{Bi_{SC}}(k) = KoR(i) \times E_{PV}(k)$$
$$\text{if } E_{Bi_{SC}}(k) > E_{Bi_C}(k) \text{ then } E_{Bi_{SC}}(k) = E_{Bi_C}(k) \quad (9)$$

The calculation principle of the static allocation for the second scenario was the same.

Concerning the customised dynamic sharing, its rules were defined according to the highest economic benefits. Taking into account the above explained economic conditions, for the energy produced and consumed at ESTIA1, neither the grid tariff nor the taxes were considered. For the PV energy consumed in ESTIA2, the grid tariff was applied but not the taxes. Finally, regarding ESTIA4 PV consumption, the grid tariff and the taxes were applied. Therefore, the defined rules were as follows (see flow-chart of Fig. 5):

- The PV generation was first associated to ESTIA1, up to its maximum consumption.
- The surplus was associated with ESTIA2, up to its maximum consumption
- The surplus of ESTIA2 was associated with ESTIA4, up to its maximum consumption.
- The surplus of ESTIA4 was fed back into the grid to be sold to the French electricity generation company that buys the PV electricity, EDF OA.

## 4. Results and discussions

This chapter presents the main results obtained in this research study. Tests were carried out in the two scenario mentioned to compare the three possible types of sharing: static, dynamic by default and dynamic. The results obtained in both scenarios are analysed in the following sub-chapters.

**Table 3**
Defined KoRs for the static sharing between the 3 buildings in each scenario.

|  | KoR (%) | |
| --- | --- | --- |
| Building | First scenario | Second scenario |
| ESTIA1 | 42.45 % | 9.44 % |
| ESTIA2 | 50.39 % | 11.20 % |
| ESTIA4 | 7.16 % | 79.36 % |
| Data Centre | | |

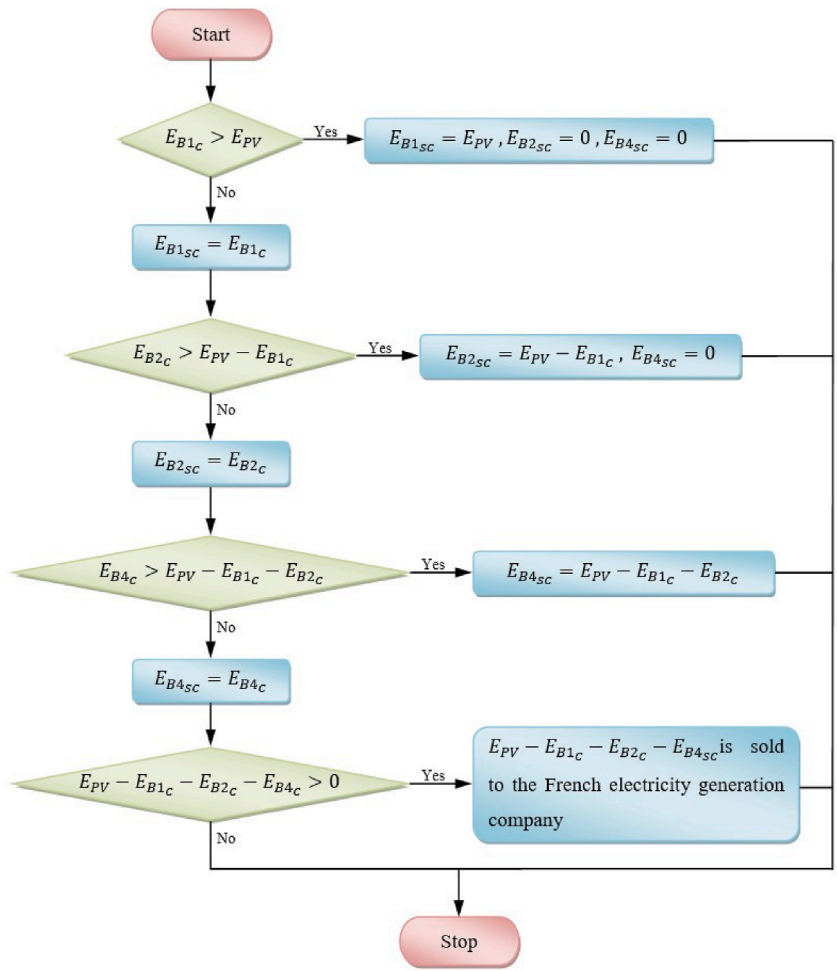

**Fig. 5.** Flow-chart representing the customized dynamic allocation.

### 4.1. Situation during the EKATE project

In order to make the analysis of the different shares as complete as possible, two representative cases related to PV production were considered: a case with a cloudy day, low solar radiation (4 May 2022) and a second one with a sunny day, high radiation level (9 May 2022).

#### 4.1.1. Low solar radiation

Fig. 6 shows the results of this case. Red, green and light purple show consumption at ESTIA1, 2 and 4 respectively. The blue curve represents the PV production at ESTIA1 (taking into account the gain mentioned above). The dark red, green and purple parts show the allocation of PV generation at ESTIA1, 2 and 4 respectively.

It can be seen that the static allocation did not allow all consumption to be counted as self-consumption. The SCR was 85.1 %. Part of the production that could have been self-consumed was not. Thus, depending on the weather and the consumption in each building, the rigidity of the approach leaded to a non-optimal SCR and then to economic losses.

The dynamic default allocation clearly solved this problem. The obtained SCR was 88.4 %. Thanks to the allocation made in proportion to the consumption of each building, the amount of PV sold energy was lower than in the static allocation. Indeed, with dynamic allocation, the consumption of the three buildings involved in the CSC was harnessed, avoiding possible allocation saturations related to the consumption in individual buildings. Equations (8) and (9) allows understanding this fact better.

As for the dynamic customised allocation, the SCR was also 88.4 %. However, thanks to the rules that were defined, it is clear that priority was given to the consumption of ESTIA1, then ESTIA2 and then ESTIA4. Thus, by comparing the results obtained in this case with those of the other two types of energy allocation, it can be seen that as long as the PV production was lower than the consumption at ESTIA1, all the production was associated with the consumption of this building. This sharing allowed the CSC system considering better the economic point of view, since, as mentioned before, no grid tariff and taxes were applied to the consumption in ESTIA1.

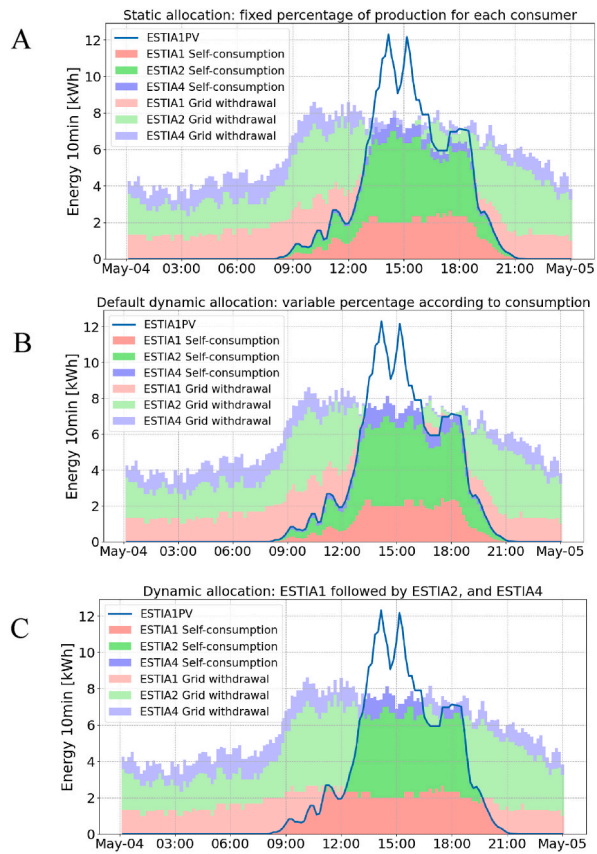

**Fig. 6.** Results of the energy allocation in the first scenario with low solar radiation level. (A) Static allocation: fixed percentage of production for each consumer. (B) Default dynamic allocation: variable percentage according to consumption. (C) Dynamic allocation: ESTIA1 followed by ESTIA2, and ESTIA4.

*4.1.2. High solar radiation*

Fig. 7 illustrates the results of the first scenario in which the data centre was not involved, but unlike Fig. 6, the results correspond to a day with a high level of radiation. Thus, it can be seen that in all three cases of sharing, the production was much higher than the cumulative consumption at mid-day.

As before, it is observed that with the static sharing, when the PV production was not very high, a small part of it was not considered as self-consumed. The obtained SCR was 48.9 %. Part of the PV production that could be self-consumed was not. This surplus production was sold at a low rate and part of the consumption was bought at a higher rate, leading to a disappointing economic result.

As with low radiation, thanks to the dynamic default sharing, all PV production was self-consumed, at least when consumption was higher than production, of course. The SCR was 49.2 %, slightly higher than with the static allocation.

As for the customised dynamic allocation, the SCR was also 49.2 %. Nevertheless, thanks to the rules defined in 3.4, it is observed again that priority was given to the consumption of ESTIA1, then to ESTIA2 and then to ESTIA4, thus managing sharing more effectively from an economic point of view.

*4.2. With the data centre*

In this second scenario, as above, the results obtained in two days with two different levels of sun radiation were compared.

*4.2.1. Low solar radiation*

Fig. 8 illustrates the results obtained in this scenario on a day with little sun. It can immediately be seen that the overall consumption was much higher thanks to the inclusion of the data centre in the CSC.

In the case of the static allocation, in comparison with the two previous cases, it can be seen that there were no losses in terms of self-consumption. The SCR was 100 %. This result is obviously due to a much higher consumption than production at every moment of the day.

For the default dynamic allocation case, the SCR was also 100 %. Furthermore, it can be noted that the allocation between the three buildings was similar to the static allocation case, with a slightly higher share for ESTIA2.

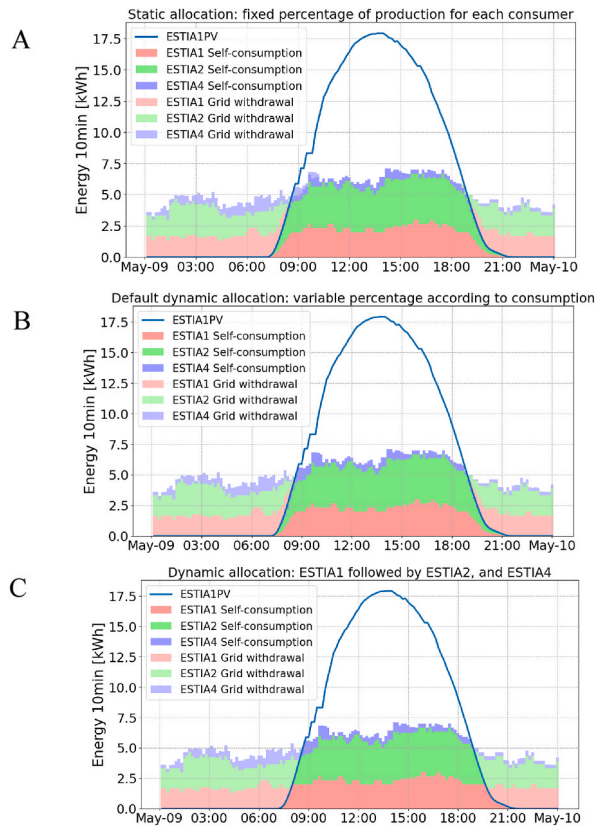

**Fig. 7.** Results of the energy allocation in the first scenario with high solar radiation level. (A) Static allocation: fixed percentage of production for each consumer. (B) Default dynamic allocation: variable percentage according to consumption. (C) Dynamic allocation: ESTIA1 followed by ESTIA2, and ESTIA4.

With regard to the case of customised dynamic sharing, the SCR was also 100 %. As before, the effect of the rules giving priority to consumption by ESTIA1, then ESTIA2, and then ESTIA4/data centre can last is clearly visible.

*4.2.2. High solar radiation*
In this scenario (Fig. 9), overall consumption was still greater than production, but the difference was smaller.
The obtained SCR was 100 % with all types of sharing.
As in the previous sections, dynamic sharing by default leaded to an allocation very similar to static sharing.
Finally, again, the personalised dynamic energy allocation made it possible to better manage the economic criterion of the CSC, by giving priority to the consumption at ESTIA1 and then at ESTIA2.

*4.2.3. Synthesis*
Fig. 10 summarizes the SCR obtained in each scenario analysed in the precedent section. In addition, in order to analyse better the sensitivity of the KoR on the SCR and the savings, a new static allocation based on the PV panels' investment was introduced. In the reality, ESTIA was the only investor in the Izarbel CSC. Thus, a fictitious scenario that assumed a different owner for each building and equal investment by each owner was created (Static33 case in Fig. 10).

Concerning the obtained results analysis, on the one hand, it can be observed that the SCR was higher with low insolation. On the other hand, it was higher when the consumption of the data centre was considered, that is, when the consumption increased. Moreover, it can be observed that the customised dynamic and default dynamic sharing types helped to maximize the SCR. Both types of allocation achieved the highest possible SCR. With regard to the Static33 investment-based allocation strategy, it was the strategy that leads to the worst SCR.

Fig. 11 shows the achieved savings in each scenario. The savings were clearly greater with strong sunlight. Regarding the influence of whether or not the data centre was included in the PMO, it is a bit curious to see that the savings were higher with the default static (for high radiation only) and dynamic sharing types in the scenario where the data centre was considered. Indeed, in both types of allocation, with the data centre, the energy allocated to ESTIA4 was significantly higher, while that allocated to ESTIA1 and ESTIA2 was lower, even though these two buildings were the most interesting in terms of savings. The reason for this result is the much higher level of self-consumed energy when the data centre was considered.

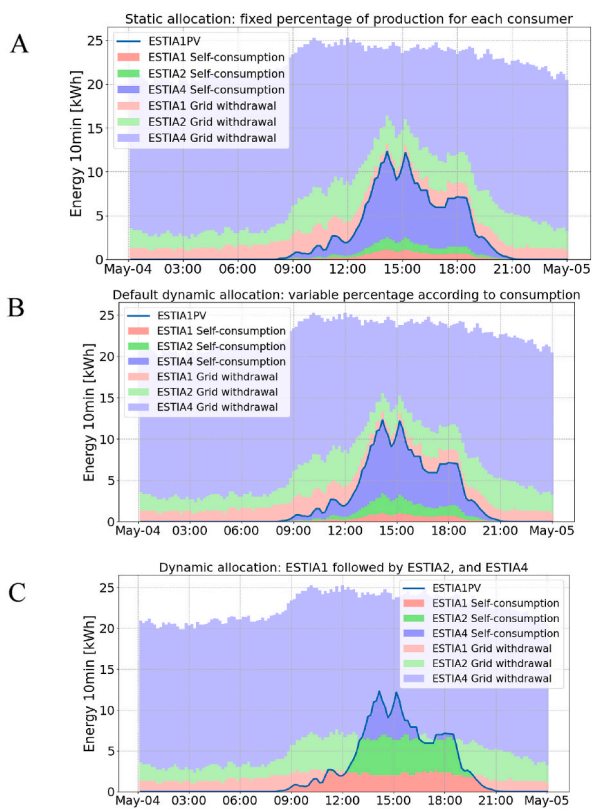

**Fig. 8.** Energy allocation results in the second scenario with low solar radiation level. (A) Static allocation: fixed percentage of production for each consumer. (B) Default dynamic allocation: variable percentage according to consumption. (C) Dynamic allocation: ESTIA1 followed by ESTIA2, and ESTIA4.

Another curiosity is that higher savings were obtained with the Static33 allocation when data centre was considered, comparing with the other static and the default dynamic allocations. This is due to the fact that with the Static33 KoRs, more self-consumed energy was allocated to ESTIA1 and ESTIA2 buildings (33 % for each).

When comparing the savings obtained with the customised dynamic sharing and the default dynamic sharing, the advantage of the customised dynamic allocation becomes clear. The lowest difference in savings corresponded to the case without the data centre.

## 5. Conclusion

The research study presented in this paper analysed the different types of PV energy sharing between consumers in the real Izarbel CSC case study, from the point of view of the SCR and the economic savings.

Several sharing scenarios made it possible to compare the three types of allocation that can be used under the French regulatory framework. From the point of view of the SCR, the interest of a dynamic sharing must be underlined, whether it is customised dynamic sharing or default dynamic sharing. However, when the static allocation is well fitted, as explained in section 3.4, the difference between static and dynamic sharing is relatively small. Indeed, compared with the static type of sharing based on consumption, the SCR of dynamic allocations was 3.86 % higher in the first scenario (low solar radiation and no data centre), 0.78 % higher in the second scenario (high solar radiation and no data centre), and 0 % higher in the third and fourth scenarios (low solar radiation and data centre and high solar radiation and data centre). The static sharing based on investment was clearly the worst-performing allocation strategy in terms of SCR.

From an economic savings perspective, the differences between the three sharing types were higher in general. In the first scenario (no DC and low solar radiation), the customised dynamic sharing got an increase of 4.84 % over the consumption-based static allocation and 4.25 % over the dynamic default. In the second scenario (no DC and high solar radiation), an increase of 0.67 % was obtained compared with the static sharing and 0.64 % compared with the dynamic sharing by default. In the third scenario (DC and low solar radiation), an increase of 41.50 % was achieved compared with the static sharing and 36.65 % compared with the dynamic sharing by default. Finally, in the fourth scenario (DC and high solar radiation), the dynamic sharing achieved 16.29 % more savings compared with the static sharing and 13.26 % compared with the default dynamic sharing. Regarding the investment-based static allocation, comparing with the consumption-based static allocation and the default dynamic allocation, the fact that more self-consumed energy was associated to ESTIA1 and ESTIA2 buildings allowed compensating the lower SCR and obtaining better

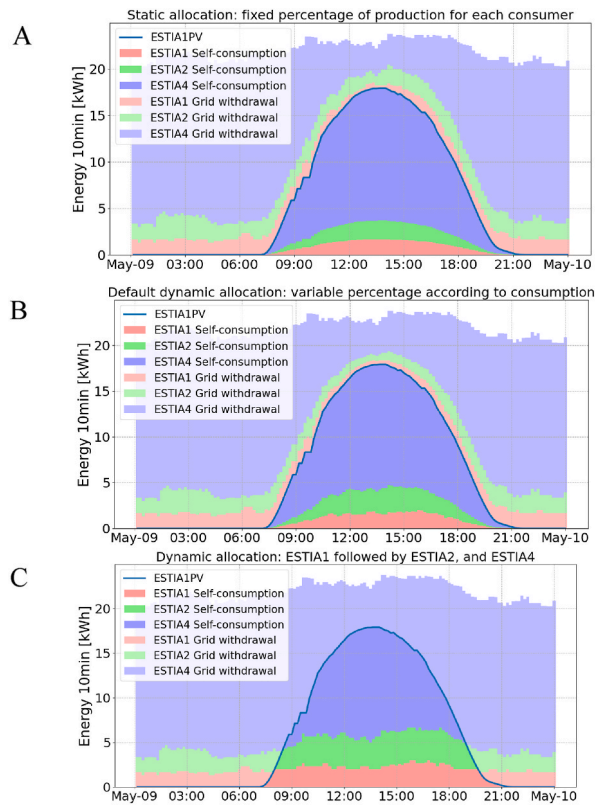

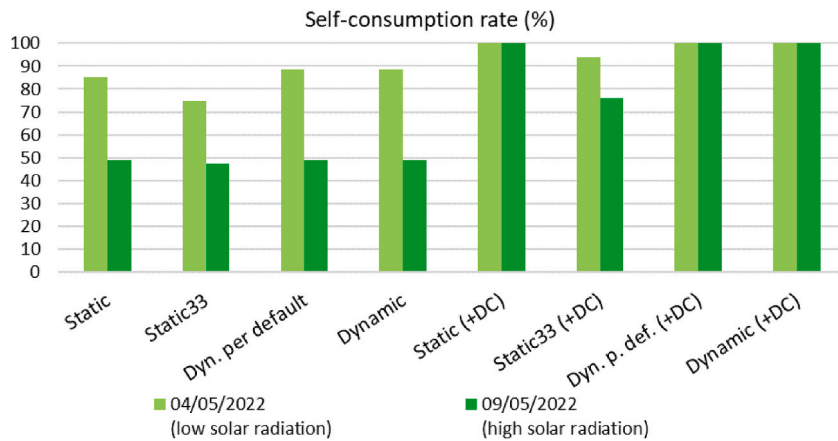

**Fig. 9.** Energy allocation results in the second scenario with a high solar radiation level. (A) Static allocation: fixed percentage of production for each consumer. (B) Default dynamic allocation: variable percentage according to consumption. (C) Dynamic allocation: ESTIA1 followed by ESTIA2, and ESTIA4.

**Fig. 10.** Synthesis of the SCR obtained in each scenario, in %.

savings when data centre of ESTIA4 was considered.

These analyses clearly show the importance of correctly selecting the members of the PMO, as the data centre in this case. I addition, it is worth highlighting the obvious advantage of using customised dynamic sharing.

This study has some limitations of course. It was carried out on the basis of two special days only. In future work, the study should be generalised by analysing data over a longer period, such as a full year. Another limitation is the need to emulate the production of the PV installation of 149 kWp. Once the new photovoltaic panels have been installed, further tests should be carried out to validate the results obtained in this study. In addition, it would be interesting to analyse the results obtained with sharing keys designed with other criteria. As well, it would be very motivating to study the use of an optimization function in the customised dynamic sharing in a more complex CSC.

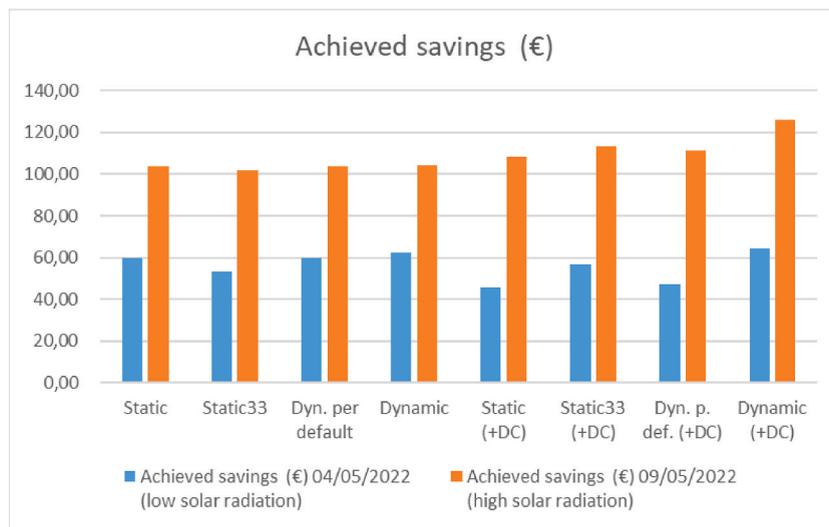

**Fig. 11.** Synthesis of the achieved savings obtained in each scenario, in €

Concerning other future studies, the project consortium plans to continue working on this topic, expanding the geographical and thematic scope of EKATE. In this new project, the scope of the CSC would be extended to the entire technology park. An important objective would be to optimise electricity sharing by taking into account the different electricity purchase and sale tariffs and a new KoR proposed by Enedis, which will allow from the end of 2022 personalised dynamic unbundled sharing, i.e. customised dynamic sharing for each PV installation.

**Data availability statement**

Data associated with this study will be made available on request.

**CRediT authorship contribution statement**

**Haritza Camblong:** Conceptualization, Data curation, Funding acquisition, Project administration, Supervision, Writing – original draft. **Octavian Curea:** Conceptualization, Data curation, Funding acquisition, Software. **Juanjo Ugartemendia:** Formal analysis, Project administration, Writing – review & editing. **Zina Boussaada:** Data curation, Investigation, Software. **Iban Lizarralde:** Conceptualization, Funding acquisition, Project administration, Supervision. **Garazi Etxegarai:** Formal analysis, Investigation, Validation.

**Declaration of competing interest**

The authors declare that they have no known competing financial interests or personal relationships that could have appeared to influence the work reported in this paper.

**Acknowledgements**

This project was co-financed up to 65% by the European Regional Development Fund (ERDF) under the Interreg V-A Spain-France-Andorra Program (POCTEFA 2014-2020, grant number EFA312/19).
Likewise, we would like to thank the company Tecsol for its contribution to the Izarbel pilot project, especially for sharing its expertise on blockchain and the French regulatory framework on energy sharing in CSC.

H. Camblong et al.                                                                                                                                                                                                                                                                                                                                                                                                                                                                                                                                                              Heliyon 9 (2023) e22252

15